\documentclass[%
amsmath,amssymb,floatfix,pra,notitlepage,superscriptaddress,longbibliography,nofootinbib
]{revtex4-1}

\usepackage{hyperref}
\usepackage[usenames,dvipsnames]{color}
\usepackage{bm}
\usepackage{graphicx}

\makeatletter
\newcommand{\manuallabel}[2]{\def\@currentlabel{#2}\label{#1}}
\makeatother

\newcommand{\ket}[1]{|#1\rangle}
\newcommand{\bra}[1]{\langle #1|}
\newcommand{\muB}{\mu_\mathrm{B}}
\newcommand{\hel}{\hat{\mathcal{H}}_{\mbox{\scriptsize el}}}
\newcommand{\hfs}{\hat{\mathcal{H}}_{\mbox{\scriptsize FS}}}
\newcommand{\hhfs}{\hat{\mathcal{H}}_{\mbox{\scriptsize HF}}}
\newcommand{\hmd}{\hat{\mathcal{H}}_{\mbox{\scriptsize Z}}}
\newcommand{\hint}{\hat{\mathcal{H}}_{\mbox{\scriptsize I}}}
\newcommand{\hamio}{\hat{\mathcal{H}}_0}

\manuallabel{chapterTormaSengstock}{1}
\manuallabel{chapterKollath}{3}
\manuallabel{chapterKokkelmans}{4}
\manuallabel{hyperfine}{4.1} 
\manuallabel{sec:ultracoldKokkelmans}{4.5} 
\manuallabel{chapterSengstock}{5}
\manuallabel{chapterChin}{6}
\manuallabel{chapterWeitenberg}{7}
\manuallabel{chapterFolling}{8}
\manuallabel{chapterParish}{9}
\manuallabel{chapterTorma}{10}
\manuallabel{Rabi}{10.2}
\manuallabel{chapterTarruell}{11}
\manuallabel{chapterKohl}{12}
\manuallabel{chapterSantos}{13}
\manuallabel{chapterPfau}{14}
\manuallabel{pfaudipolartable}{14.1} 

\begin{abstract}
We provide an introduction to the experimental physics of quantum gases. At the low densities of ultracold quantum gases, confinement can be understood from single-particle physics, and interactions can be understood from two-body physics. The structure of atoms provides resonances both in the optical domain and in the radio-frequency domain. Atomic structure data is given for the 27 atomic isotopes that had been brought to quantum degeneracy at the time this chapter was written. We discuss the motivations behind choosing among these species. We review how static and oscillatory fields are treated mathematically. An electric dipole moment can be induced in a neutral atom, and is the basis for optical manipulation as well as short-range interactions. Many atoms have permanent magnetic dipole moments, which can be used for trapping or long-range interactions. The Toronto $^{40}$K/$^{87}$Rb lattice experiment provides an illustration of how these tools are combined to create an ultracold, quantum-degenerate gas.
\end{abstract}

\begin{document}
\title{Making an Ultracold Gas \label{chapterThywissen}}
\author{Dylan Jervis}
\affiliation{Department of Physics, University of Toronto, M5R 2K7 Canada}
\author{Joseph H. Thywissen}
\affiliation{Department of Physics, University of Toronto, M5R 2K7 Canada}
\affiliation{Canadian Institute for Advanced Research, Toronto, M5G 1Z8 Canada}
\maketitle

\tableofcontents


\section{Introduction: Quantum gases must be dilute and ultracold \label{sec:introYYZ} }

The quantum gases discussed in this text are ``quantum'' in the many-body sense: the spacing between particles $n^{-1/3}$ is comparable or less than their thermal de Broglie wavelength 
$\lambda_{T} = \sqrt{{2 \pi \hbar^2}/{M k_B T} },$ 
where $n$ is the number density, $M$ is the particle mass, and $T$ is the temperature. This is typically achieved at $n \approx 10^{19}$\,m$^{-3}$, such that particles are hundreds of nanometers apart. For $^{87}$Rb (the first \cite{Cornell:87Rb} and still most common gas Bose condensed), $\lambda_T\sim100$\,nm requires a temperature of roughly 100\,nK. Achieving such a low temperature was a tremendous technical challenge, and not realized until 70 years after Einstein first wrote down the criterion for Bose condensation \cite{Einstein,Cornell:2002vo,Ketterle:2002to}. But why did experimentalists not make their lives easier, and work at both a higher density and higher temperature?

For instance, at the density of air, about $2.5 \times 10^{25}$\,m$^{-3}$, one could reach quantum degeneracy at a temperature $T\approx 2$\,mK. Even more optimistic: at fixed {\it pressure}, cooling an ideal gas increases its density, so cooling a sample of argon at fixed atmospheric pressure could achieve quantum degeneracy at 0.5\,K and $n \sim 10^{28}$\,m$^{-3}$. Unfortunately, if you were to try this experiment, for instance with a dilution refrigerator, you would find that the gas simply freezes.

In equilibrium, a nanokelvin sample of $^{87}$Rb is a solid, not a gas \cite{Cornell:2002vo}. However, {\it quantum gases are a metastable phase of matter.} Measurements must be performed before the gas realizes that it ``should be'' a solid at nanokelvin temperatures. The lifetime of this metastable condition is given by the rate of three-body loss, the process by which three atoms are converted to a bound dimer and a free atom that carries away the binding energy. The rate of this loss process is $L\,n^2$, where $L \approx 10^{-40}$\,m$^6$/s for $^{87}$Rb, for instance \cite{Burt:1997uy,Soding:1999vr}. For a lifetime of 1\,s, one requires $n \lesssim 10^{20}$\,m$^{-3}$. This is five to eight orders of magnitude more dilute than the quantum gas achieved by the naive approaches described above. Instead, one is forced to work at extremely low density.

While inelastic scattering gives an upper bound on practical density, {\it elastic} scattering bounds density from below. The elastic cross section between neutral atoms is roughly $\sigma=10^{-15}$\,m$^2$. The collision rate for such a gas near quantum degeneracy is roughly 100\,s$^{-1}$ when the density is $10^{19}$\,m$^{-3}$. This rate would allow for many elastic collisions during (for instance) a one-second-long experiment, enabling a thermal equilibrium between the translational degrees of freedom. Although it is understood that the molecular degrees of freedom are not described by a temperature (due to the slow relaxation rate at low density), we will describe the centre-of-mass position and velocity of the quantum gas by a temperature $T$.

Lower densities, however, will reduce the thermalization rate of the gas in proportion to the density. The collision rate is $\gamma = n \sigma v_{T}$, where $v_{T} = \sqrt{8 k_B T/\pi M}$ is the relative thermal velocity \cite{Walraven:1996tk,KetterleVanDruten:evap}. Constraining the temperature to the quantum regime, $n \lambda_{T}^3 = 1$, the collision rate scales as $n^{4/3}$. So if instead of $n\sim10^{19}$\,m$^{-3}$ we had chosen a density a hundred times smaller, the collisional rate would be less than one per second. Eventually one is limited by the finite lifetime of trapped atoms in a vacuum system, and the finite lifetime of the experimentalist.

These two considerations - fast translational equilibrium yet slow three-body loss - constrains quantum gas experiments to work with dilute samples. There is some variation between elements, but even comparing hydrogen $^{1}$H ($n \approx 2 \times 10^{20}$\,m$^{-3}$ in  Ref.~\onlinecite{Fried:1H}) and cesium $^{133}$Cs ($n \approx 1 \times 10^{19}$\,m$^{-3}$ in Ref.~\onlinecite{Grimm:133Cs}), the range is not very large.
It is interesting to note that putting a single atom in each site of a cubic optical lattice with a typical period of $a=0.5\,\mu$m gives a density of roughly $n \sim 10^{19}$\,m$^{-3}$ (Chapter~\ref{chapterKollath}, Chapter~\ref{chapterSengstock}).

The density sets a typical energy scale, $\epsilon = \hbar^2 n^{2/3} / M$, at which many-body physics of interest might occur. For weakly interacting bosons, the critical temperature for superfluidity \cite{Pethick,Ketterle1999} occurs at $k_B T_c \approx 3.3 \epsilon$. For low-temperature fermions, the Fermi energy \cite{Pethick,CastinFermi} is $E_F \approx 7.6 \epsilon$. The transition to a superfluid of paired fermions occurs at $0.17\,E_F$ for a unitary Fermi gas  \cite{Ketterle2008,Ku:2012ue}, and lower for a weakly attractive gas (Chapter~\ref{chapterParish}).
For a density of $10^{19}$\,m$^{-3}$ and mass 87\,amu, $\epsilon/k_B$ is 25\,nK, hence the moniker ``ultracold atoms'' for this field of research.

In the remainder of this chapter, we describe several tools common to achieving quantum degeneracy in gases. 
In Section~\ref{sec:atomsYYZ} we discuss the ground-state structure of atoms in static fields, which enables magnetic trapping and Stern-Gerlach separation.
Then in Section~\ref{sec:oscfieldsYYZ}, we describe the response to a driving field, which enables optical imaging (Chapter~\ref{chapterChin}, Chapter~\ref{chapterWeitenberg}, Chapter~\ref{chapterFolling}), laser cooling, optical trapping, and RF spectroscopies (Chapter~\ref{chapterTorma}, Chapter~\ref{chapterTarruell}).
Interactions are what makes many-body states interesting, and are introduced in Section~\ref{sec:interactionsYYZ}, but discussed in more detail in Chapter~\ref{chapterKokkelmans}, Chapter~\ref{chapterSantos}, and Chapter~\ref{chapterPfau}. 
We conclude in Section~\ref{sec:sequenceYYZ} with a description of the experimental sequence used to produce a quantum gas of $^{87}$Rb and $^{40}$K in the Toronto lattice experiment, as a specific example of how these tools are combined in experiments all over the world.


\section{Atomic structure and response to static fields \label{sec:atomsYYZ} }

At the low density of quantum gas experiments, manipulation of atoms can be understood with single-particle and two-particle pictures. In this section we describe the ground-state structure of an isolated atom, and how it is perturbed by static electric and magnetic fields. The ground-state structure explains why the rare-earth fermions are uniquely well suited for SU(N) physics  \cite{Hofstetter:2004,Gorshkov:2010ua}, for instance. Magnetic properties highlight the interest of chromium and other transition-metal atoms. Electric properties are especially important for dipolar molecules  \cite{Carr:2009wz}, and describe the low-frequency asymptote of the optical trapping potential of atoms.

\subsection{Ground state structure of quantum gases \label{sec:structureYYZ} }

In the absence of external fields, there are three terms in the Hamiltonian of hydrogen-like atoms,
\begin{equation} \label{eq:H0YYZ}
\hamio=\hel+\hfs+\hhfs,
\end{equation}
where $\hel$ contains the non-relativistic kinetic energy of the electrons and the Coulomb interaction between them; $\hfs$ is the fine structure term that includes relativistic corrections to $\hel$, electron spin, and spin-orbit terms; and $\hhfs$ is the hyperfine structure term induced by the spin and electric quadrupole moment of the nucleus.

The ground states, as well as other properties, of the 27 isotopes that have been cooled to quantum degeneracy%
 \cite{Fried:1H,Jaakkola:1H,Vassen:3He,Aspect:4He,Cohen:4He,Hulet:6Li,Salomon:6Li,Bradley:1995vn,Sackett:1997wr,Ketterle:23Na,Roati:39K,Jin:40K,%
Modugno:41K,Sterr:40Ca,Pfau:52Cr,Schreck:84Sr,Killian:84Sr,Schreck:86Sr,Killian:87Sr,Killian:88Sr,Wieman:85Rb,Cornell:87Rb,Grimm:133Cs,%
Takahashi:168Yb,Takahashi:170Yb,Takahashi:171Yb,Takahashi:173Yb,Takahashi:174Yb,Takahashi:176Yb,Lev:161Dy,Lev:164Dy,Ferlaino:168Er}
 are given in Table~\ref{tab:atoms}. The Russell Saunders notation gives the total angular momenta in the form $^{2s+1}l_{j}$,  using the letters \{S, P, D, F, G,  \ldots \} to indicate orbital angular momentum $l=\{0,1,2,3,4, \ldots\}$, but numbers for spin $S$ and total electronic angular momentum $j$. As an example, all Group~I atoms have a $^2 \mathrm{S}_{1/2}$ ground state, meaning $s=1/2$, $l=0$, and $j=1/2$. Atoms beyond Group~I have more than one electron in the outer shell. For all atoms listed, the inner electronic configuration is given using another standard notation: $n \ell^{k}$, where $n$ is the principal quantum number, $\ell$ is again the orbital angular momentum given as a letter, and $k$ is the number of electrons in that shell.

For atoms with non-zero nuclear and electronic spin, $\hhfs$ splits the ground states. This splitting, $\Delta E_\mathrm{HF}$, is given for the Group~I atoms where there are only two levels at zero field. The combined total angular momentum is called $f$. For example, for $^{87}$Rb the nuclear spin is $i=3/2$, which, added to the ground-state electronic angular momentum of $1/2$ (all of it from spin), gives either $f=1$ or $f=2$. Unlike electronic excited states, there is no significant spontaneous decay of the upper hyperfine states, so these are often called ground states. Nuclear structure is also important for control of scattering properties of atoms, using Fano Feshbach resonances, discussed in Chapter~\ref{chapterKokkelmans} and leading to the physics discussed in Chapter~\ref{chapterParish}.

Alkali atoms (Li, Na, K, Rb, Cs) were the first to be cooled to quantum degeneracy and remain the most widespread in use, because their response to optical excitation is simple. The lowest-lying excited state is $n^2\mathrm{P}$, which has the same principle quantum number as the $n\mathrm{S}$ ground state. The transition energy (roughly 1\,eV) comes from reduced electron-electron interaction in the P states, since the orbital has a larger average radius and samples less of the charge density of the core. The alkali excited state is further split into the $n^2\mathrm{P}_{1/2}$ and $n^2\mathrm{P}_{3/2}$ states by $\hfs$, with energy splittings given in Table~\ref{tab:atoms}. Atoms with a large fine-structure splitting are good candidates for spin-dependent optical potentials, discussed in Chapter~\ref{chapterSengstock}.

The splitting between the S and P states is absent in hydrogen, which pushes the first optical excitation to a much higher $n$ state, at a practically inaccessible 121\,nm wavelength. Bose condensation of hydrogen \cite{Fried:1H,Jaakkola:1H} had to proceed without laser cooling, which is why it is rarely used today. However, its historical role was significant: magnetic trapping and evaporative cooling were pioneered using hydrogen \cite{Evap:1988,Walraven:1996tk,Kleppner99}.

Atoms beyond Group~I that have been cooled to degeneracy (Table~\ref{tab:atoms}) have several remarkable features. Metastable helium (He*) is amenable to single-atom detection at a microchannel plate, since their internal energy (20\,eV) is sufficient to overcome the work function of a metal and extract an electron. Atoms with closed outer shells (Ca, Sr, Yb, Er, Yb) have narrow optical lines for laser cooling, clock transitions, and optical Feshbach resonances. Finally, atoms with higher angular momentum can have stronger magnetic dipole moments: Cr, Er, and Dy have ground states with 6$\muB$, 7$\muB$, and 10$\muB$, respectively; $\muB$ is the Bohr magneton. We discuss the effect of an external magnetic field on a single atom in the next section, but the real excitement about these high-$\mu$ atoms comes from the dipole-dipole interactions discussed in Section~\ref{sec:interactionsYYZ},  Chapter~\ref{chapterSantos} and Chapter~\ref{chapterPfau}.

\begin{table}[!t]
\caption{Zoology of cold: Atomic data for isotopes that have been cooled to quantum degeneracy. While the Group I atoms are the most commonly used and serve as the reference for the field, this table indicates several remarkable features of other elements: magnetic moments, metastability, and narrow transitions. Fermionic isotopes are those with integer nuclear spin $i$ for the Group I elements, and half-integer $i$ for the atoms beyond Group I.}
{\begin{tabular}{@{}clccccc@{}} \toprule
\multicolumn{7}{l}{{\bf Group I elements:}}\\ 
Element& Ground state& $\Delta E_\mathrm{FS}/h$ &  Cooling lines   & Isotope & i &$\Delta E_\mathrm{HF}/h$  \\
             &  &  (THz)  & $\Gamma/h$ \& $\lambda$ & & & (MHz) \\ \colrule
Hydrogen 	& 1s$^1$ $^2$S$_{1/2}$ 	&	0.02	& 	{\em (not laser cooled)}  & $^1$H 	&  1/2 	&1420 \\ 
Lithium 		& 1s$^2$2s$^1$ $^2$S$_{1/2}$ 	& 0.01  	& 5.9\,MHz at 671\,nm 	& $^6$Li 	& 1 		& 228 \\
    			&      	&  		&	           			  754\,kHz at 323\,nm & $^7$Li   & 1/2 	& 803.5  \\ 
Sodium 		& [Ne]3s$^1$ $^2$S$_{1/2}$ 	& 0.52  	& 9.8\,MHz at 589\,nm 	& $^{23}$Na &  3/2   & 1771.6 \\ 
Potassium 	& [Ar]4s$^1$ $^2$S$_{1/2}$ 	&  1.73  	&  6.0\,MHz at 767\,nm 	& $^{39}$K & 3/2 	& 461.7   \\
 			&   		&   		&  			1.2\,MHz at 405\,nm 		& $^{40}$K &  4  &  1285.8 \\
 			&   		&   		&  					& $^{41}$K & 3/2 & 254.0 \\ 
Rubidium 		& [Kr]5s$^1$ $^2$S$_{1/2}$    	& 7.12  	& 6.1\,MHz at 780\,nm 	& $^{85}$Rb  & 5/2 & 3035.7 \\
 			&   		&   		&  					& $^{87}$Rb &  3/2 & 6834.7   \\  
Cesium  		& [Xe]6s$^1$ $^2$S$_{1/2}$  	&  16.61  & 5.2\,MHz at 852\,nm 	& $^{133}$Cs &  7/2 & 9192.6  \\  
\toprule
\multicolumn{7}{l}{{\bf Elements beyond Group I:}}\\ 
Element &  \multicolumn{2}{l}{Ground state}   & Cooling lines  & Isotope & i & $\mu^\mathrm{max}_\mathrm{gnd} / \muB$ \\ \colrule
Helium &  \multicolumn{2}{l}{2$^{3}$S$_0$ (metastable)}  & 1.6\,MHz at 1083\,nm 	& $^3$He* & 1/2 & 2 \\ 
& \multicolumn{2}{l}{} &   1.5\,MHz at 389\,nm	& $^4$He* & 0 & 2 \\ 
Calcium & \multicolumn{2}{l}{[Ar]4s$^2$ $^1$S$_0$} & 34\,MHz at 423\,nm  &  $^{40}$Ca & 0 & 0\\ 
  &  \multicolumn{2}{l}{}  & 370\,Hz at 657\,nm &  \\ 
Chromium & \multicolumn{2}{l}{[Ar]3d$^5$4s$^1$ $^7$S$_3$}    & 5\,MHz at 426\,nm &  $^{52}$Cr & 0 & 6\\ 
Strontium & \multicolumn{2}{l}{[Kr]5s$^2$ $^1$S$_0$} &  32\,MHz at 461\,nm & $^{84}$Sr & 0 & 0\\ 
 & & & 												 7.4\,kHz at 689\,nm & $^{86}$Sr & 0 & 0 \\ 
 & & & 												  & $^{87}$Sr & 9/2 & 0.00059 \\ 
 & & & 												  & $^{88}$Sr & 0 & 0 \\ 
Dysprosium & \multicolumn{2}{l}{[Xe]4f$^{10}$6s$^{2}$ $^5$I$_8$}   & 32\,MHz at 421\,nm& $^{161}$Dy & 5/2 & 10 \\
 &\multicolumn{2}{l}{}  & 1.8kHz at 741\,nm & $^{162}$Dy & 0 & 10\\ 
 & & & 												& $^{164}$Dy &  0 & 10\\ 
Erbium & \multicolumn{2}{l}{[Xe]4f$^{12}$6s$^{2}$ $^3$H$_6$} & 28\,MHz at 401\,nm & $^{168}$Er & 0 & 7\\ 
  & \multicolumn{2}{l}{} & 190\,kHz at 583\,nm  &  \\ 
  & \multicolumn{2}{l}{} & 8\,kHz at 841\,nm &  \\ 
Ytterbium & \multicolumn{2}{l}{[Xe]4f$^{14}$6s$^{2}$ $^1$S$_0$}  & 29\,MHz at 399\,nm & $^{170}$Yb  & 0 & 0\\ 
 & & & 										182\,kHz at 556\,nm &  $^{171}$Yb & 1/2  & 0.00027\\ 
 & & &  													 &  $^{173}$Yb & 5/2 & 0.00037 \\ 
 & & & 	 		 									  	 & $^{174}$Yb & 0 & 0 \\ 
 & & & 													 &  $^{176}$Yb & 0 & 0\\ 
\botrule
\end{tabular}}
\label{tab:atoms}
\end{table}

\subsection{The Zeeman effect \label{sec:ZeemanYYZ} }

Applying a magnetic field to the atom adds another term to the Hamiltonian, 
\begin{equation} \label{eq:HZYYZ}
\hmd = -\hat\mu \cdot \mathbf{B}, 
\end{equation}
where $\hat\mu$ is the magnetic dipole operator and $\mathbf{B}$ is the external field. This term breaks the rotational symmetry of the Hamiltonian of an isolated atom.  Solving for the eigenvalues of the electronic ground and excited states in an applied magnetic field is a standard exercise in an atomic physics course. For the ground state of alkali atoms, the energy eigenvalues are given by the Breit-Rabi formula,
\begin{equation} \label{eq:BR}
\frac{E}{\Delta E_\mathrm{HF}} =  -\frac{1}{4 f_+} \pm \frac{1}{2} \sqrt{ 1 + 2 \frac{m_f}{f_+} x + x^2 },
\end{equation}
where $x = g_j \muB B/\Delta E_\mathrm{HF}$, $m_f=m_i \pm 1/2$ is the magnetic quantum number (projection of total angular momentum along the field axis), $B=|\mathbf{B}|$ is the magnitude of the field, $\Delta E_\mathrm{HF}$ is the hyperfine splitting, given in Table~\ref{tab:atoms}, and $f_+=i+1/2$.%
\footnote{There is a sign ambiguity in the square root. When $m_f=f_+$, the $\pm \frac{1}{2} \sqrt{\ldots}$ term should be replaced by $+ \frac{1}{2} (1+x)$; and when $m_f = -f_+$, the same term should be replaced by $+ \frac{1}{2} (1-x)$. We have neglected the interaction of the nuclear magnetic moment with the external field in this formula (ie, taken $g_i = 0$), but still retain $i$ and $m_i$ as quantum numbers affecting the hyperfine energy and projection of the electronic spin. To calculate microwave transitions to a precision better than 1\%, these terms should be added back in: add a term $+ (m_f x)/(g_j/g_i - 1)$ to the energy, and replace $g_j$ with $(g_j - g_i)$ in $x$.}
For ground states $g_j\approx2$, but more precision can be found in atomic data tabulated elsewhere. A plot of $E$ vs. $B$  is shown in Fig.~\ref{hyperfine} for $^{87}$Rb. 

For weak magnetic fields, the magnetic dipole term $\hmd$ can be treated as a perturbation, and the Zeeman shifts are linear: $\partial E/\partial B \rightarrow g_f m_f \muB$, where $g_f$ is the Land\'e factor. For ground states of alkalis, $g_f = \pm 1/f_+$, such that the {\em stretched state} has the moment of one $\muB$. For atoms beyond Group~I, the maximum magnetic moment of the ground state is given in Table~\ref{tab:atoms}. Magnetic and magneto-optical trapping typically takes place in this linear regime.

The Breit Rabi equation also gives the Zeeman energy at higher fields. The deviation from linearity is called the quadratic Zeeman shift, and is proportional to  $(\muB B)^2 / \Delta E_\mathrm{HF}$. At very high fields, $B \gg \Delta E_\mathrm{HF} / \muB$,  the magnetic dipole term is the dominant effect. With the hyperfine interaction now treated as a perturbation, the atomic levels become increasingly well described by the quantum numbers $m_j$ and $m_i$, these being the individual spin projections of the electron and nuclear angular momenta.  Since the magnetic moment of the electron is much larger than that of the nucleus, the energy states break up into a higher-energy set of $m_j = +1/2$ (spin up) eigenstates and a lower-energy set of $m_j=-1/2$ (spin-down) eigenstates. In this regime,  $E \rightarrow m_j g_j \muB B + A_{\rm HF} m_j m_i$, where $A_{\rm HF} = \Delta E_\mathrm{HF}/f_+$ for the alkali ground states. In each $m_j$ set, level spacing is $A_{\rm HF}$.

Magnetic traps use inhomogeneous magnetic fields to create a confining potential surface. Since Maxwell's equations forbid static fields to have local maxima in free space, neutral atoms are typically trapped in weak-field-seeking states: those with $g_f m_f > 0$. The depth of a magnetic trap depends on the strongest closed-field surface: this can be several teslas for permanent or superconducting magnets, but more commonly electromagnets are used, to give fractions of teslas, able to hold atoms cooled to 10~mK. While much lower than room temperature, this is much higher than laser-cooling temperatures (discussed in Section~\ref{sec:lasercoolingYYZ}), and thus an appropriate trap for pre-cooled atoms.

Magnetic traps have several limitations. First of all, not all atoms have strong dipole moments: atoms with $^1S_0$ ground states (Ca, Sr, Yb in Table~\ref{tab:atoms}) have no unpaired electrons, and thus at most a nuclear dipole moment. Second, even with alkali atoms, the use of a magnetic field for trapping precludes the use of the field for tuning interactions (see Chapter~\ref{chapterKokkelmans}). In either of these two cases, optical traps must be used.

\subsection{Electric fields and the Stark Effect \label{sec:StarkYYZ} }

Eigenstates of $\hel$ are symmetric under parity, and thus do not have an electric dipole moment. However an external electric field can break this symmetry and induce a dipole moment. In a semiclassical treatment, the additional Hamiltonian term is
\begin{equation} \label{eq:HIntYYZ} \hint = - \hat{\mathbf{D}} \cdot {\mathcal E} \mathbf{e}_p, \end{equation}
where $\hat{\mathbf{D}}$ is the dipole moment operator, ${\mathcal E}$ is the electric field strength, and $\mathbf{e}_p$ is the direction of the field. The induced moment interacts with the field to produce a second-order shift called the Stark Effect:
\begin{equation} \Delta E_\mathrm{Stark} = -\frac{1}{2}\alpha_0 {\cal E}^2, \label{eq:StarkYYZ} \end{equation}
where $\alpha_0$ is the dc polarizability and ${\cal E}$ is the electric field strength. Alkali atoms have $\alpha_0 \approx 3 \times 10^{-39}$\,Cm$^2$/V. Since fields greater than $10^5$\,V/m typically cause electrode discharge, static potentials cannot hold atoms hotter than a few $\mu$K, much weaker than magnetostatic traps. Furthermore, because the dipole is induced, there are no weak-field-seeking states. Since field maxima are not allowed by Maxwell's equations, electrostatic potentials must be combined with another type of potential to form a stable trap for neutral atoms.

Polar molecules, however, can have a much stronger reaction to electric fields \cite{Carr:2009wz}. For instance, RbK in its singlet ground state was measured to have an electric dipole moment of 0.57 Debye \cite{Ye:2008}, where one Debye is $3.34 \times 10^{-30}$\,Cm. In an electric field of $10^4$\,V/m, the energetic effect is roughly $10^5$ stronger than it would be for a neutral alkaline atom. Furthermore, the long-range and anisotropic interactions between oriented polar molecules lead to interesting dipolar physics. For further discussion and references, see Chapter~\ref{chapterSantos} and Chapter~\ref{chapterPfau}.


\section{Atoms in oscillating fields  \label{sec:oscfieldsYYZ} }

In many-body theory, the same low-energy Hamiltonian could often describe either electrons or atoms: for example, the Hubbard models discussed in Chapter~\ref{chapterKollath}. However unlike electrons, atoms and molecules have rich internal structure. Oscillating electromagnetic fields are the primary tool to access this structure. In this section we lay out the basic formalism of a driven two-level quantum system. Although we pursue the example of an optical field driving an electronic transition, the same concepts apply to an RF field driving a magnetic transition. Resonance and resonant enhancement underlie laser cooling (see Section~\ref{sec:lasercoolingYYZ}), optical lattices (Chapter~\ref{chapterKollath}, Chapter~\ref{chapterSengstock}), optical imaging (Chapter~\ref{chapterChin}, Chapter~\ref{chapterFolling}), spectroscopy (Chapter~\ref{chapterTorma}, Chapter~\ref{chapterTarruell}).

\subsection{The rotating wave approximation \label{sec:RWAYYZ}}

Let us consider a single resonance of $\hamio$ (Eq.~\ref{eq:H0YYZ}), in which two quantum levels, $\ket{g}$ and $\ket{e}$, are coupled by the electric dipole Hamiltonian $\hint$ (Eq.~\ref{eq:HIntYYZ}). The electric field is ${\mathcal E}(\mathbf{r},t) = {\mathcal E}(\mathbf{r}) \mathbf{e}_p \cos{(\omega_L t)}$, where $\mathbf{r}$ is the center-of-mass position of the atom, $\mathbf{e}_p$ is the polarization, and $\omega_L$ is the frequency. The coupling strength of the field is given by a matrix element 
\begin{equation} \label{eq:RabiFrequYYZ}
\hbar \Omega(\mathbf{r}) = - \langle e | \hat{\mathbf{D}} \cdot \mathbf{e}_p | g \rangle {\mathcal E}(\mathbf{r}),
\end{equation}
where $\Omega$ is the Rabi frequency. This is the rate at which the field drives population oscillation between the ground and excited states, on resonance (see Section~\ref{Rabi}).

Next, we write out the Hamiltonian in the basis $\{ \ket{g}, \ket{e} \}$. The on-diagonal terms are given by $\hamio$, and we will choose $E=0$ to be halfway between the two states. Since the electric field has odd parity, it must couple states of opposite parity, and $\hint$ is strictly off-diagonal. We are left with
\begin{align}
\hamio + \hint \longrightarrow \frac{\hbar \omega_{eg}}{2} \hat\sigma_z + \hbar \Omega \hat\sigma_x \cos{(\omega_L t)},
\end{align}
where $\hat\sigma_x = \ket{e}\bra{g} + \ket{g}\bra{e}$ and $\hat\sigma_z = \ket{e}\bra{e} - \ket{g}\bra{g}$ are Pauli matrices. If we had begun with a {\em magnetic} dipole, Eq.~\ref{eq:HZYYZ}, with an oscillating magnetic field along $x$, and a static field along $z$ (or equivalent internal energy), we would have arrived at this same equation (e.g., replacing $\omega_L$ with $\omega_{RF}$), recognizing that $\hat{S}_x \rightarrow \hbar \hat\sigma_x/2$ in the $S_z$ basis.

For a near-resonant drive, $\delta \ll \omega_{eg}$, where $\delta = \omega_L - \omega_{eg}$ is the detuning from resonance, we can make the {rotating wave approximation} (RWA), where in the rotating frame of the drive field,
\begin{equation} \label{eq:HRWAYYZ}
\hamio + \hint 
 \xrightarrow[\mathrm{RWA}]{}  \frac{\hbar}{2} (-\delta \hat\sigma_z + \Omega \hat\sigma_x ),
\end{equation}

Further discussion of this Hamiltonian, as well as references for further reading, can be found in Chapter~\ref{chapterTorma}. 

\subsection{The optical dipole potential \label{sec:dipoleYYZ} }

The prevalence of optical traps in manipulation of quantum gases is due to their versatility. Laser light can be focused with lenses, can travel through vacuum windows, can be modulated dynamically, and can interfere to create optical lattices. The trapping potential felt by a ground-state atom can be easily estimated using Eq.~\ref{eq:HRWAYYZ} in second-order perturbation theory:
\begin{equation} \label{eq:VdipRes}
V_\mathrm{dip} \approx \frac{\hbar \Omega^2}{4 \delta}.
\end{equation}
This scales as intensity divided by detuning, since $\Omega$ is proportional to field. When $\omega_L < \omega_{eg}$, the detuning is negative and atoms are attracted to high intensity. When $\omega_L > \omega_{eg}$, atoms are repelled by intensity. This potential is called the {\em dipole potential} because it follows the behavior of a driven classical oscillator.

When terms beyond the rotating-wave approximation are included, the dipole potential from a single resonance is \cite{Grimm:2000}
\begin{equation} \label{eq:VdipYYZ}
V_\mathrm{dip}(\mathbf{r}) = -\frac{3 \pi c^2}{2 \omega_{eg}^3} \left( \frac{\Gamma}{\omega_{eg} - \omega_L} + \frac{\Gamma}{\omega_{eg} + \omega_L} \right) I(\mathbf{r}),
\end{equation}
where  $I(\mathbf{r})$ is intensity, and $\Gamma$ is the line width of the transition (Section~\ref{sec:spforceYYZ}). When far from a single strong resonance, the sum over many transitions will contribute a net shift of the ground state. For light in the far infrared (for example a CO$_2$ laser at $\lambda_L\approx$10\,$\mu$m), the sum over all excited states approaches the static limit of Eq.~\ref{eq:StarkYYZ}.

Optical traps that are not in the electrostatic limit benefit from a resonant enhancement. As can be seen from the two terms of Eq.~\ref{eq:VdipYYZ}, the resonant enhancement is roughly $\omega_{eg} / \delta$. This could in principle be as large as $\omega_{eg}/\Gamma$, which is as large as $10^{12}$ for the narrow lines Table~\ref{tab:atoms}. However, in practice experimentalists use $\delta \gg \Gamma$ to avoid heating, as discussed in the next section.

\subsection{Spontaneous emission and near-resonant scattering \label{sec:spforceYYZ} }

While the formalism developed in Section~\ref{sec:RWAYYZ} is true for any kind of driven two-level quantum system, there is an important difference between RF and optical transitions: the fast rate of spontaneous emission for an electronic excited state. The decay rate of the excited state depends on the dipole matrix element:
\begin{equation} \label{eq:gammaYYZ}
\Gamma = \frac{\omega_{eg}^3}{3 \pi \epsilon_0 \hbar c^3} \sum_{p=\pi,\sigma^\pm} \left| \bra{e} \hat{\mathbf{D}} \cdot \mathbf{e}_p \ket{g} \right|^2.
\end{equation}
The cubic dependence on frequency explains the qualitative difference between RF transitions in the MHz to GHz range, compared to optical transitions that are typically 100\,THz. For various optical transitions, $\Gamma$ are tabulated in Table~\ref{tab:atoms}. If there are multiple accessible decay paths, an additional sum over those transition frequencies is needed in Eq.~\ref{eq:gammaYYZ}. However for this pedagogical treatment, let us consider {\em cycling transitions,} which connect a single excited state to the ground state, using polarization $\mathbf{e}_p$. Since both the Rabi frequency Eq.~\ref{eq:RabiFrequYYZ} and the decay rate Eq.~\ref{eq:gammaYYZ} are then proportional to the same matrix element, we can write
\begin{equation} \label{eq:IsatYYZ}
2 \frac{\Omega^2}{\Gamma^2} = \frac{I}{I_{S}}
\quad \mbox{where} \quad 
I_{S} = \frac{2 \pi^2 \hbar c \Gamma}{3 \lambda^3},
\end{equation}
and $I = c \epsilon_0 \langle {\mathcal E}^2 \rangle$, where $c$ is the speed of light, $\epsilon_0$ is the electric constant, and brackets indicate the time average.

To find the time evolution of our driven two-level system, a master equation treatment is required, since spontaneous emission is an incoherent process. The density matrix $\hat{\rho}$ evolves coherently under Hamiltonian Eq.~\ref{eq:HRWAYYZ}, but population and coherences also decay. This leads to
\begin{align} \label{eq:OBEYYZ}
\frac{d}{dt} \rho_{ee} & = -\Omega \Im \{ \rho_{eg} \} - \Gamma \rho_{ee} \\
\frac{d}{dt} \rho_{eg} & = i (\delta + i \frac{\Gamma}{2} ) + i \Omega (\rho_{ee} - \frac{1}{2}), 
\end{align}
where $\rho_{ee}$ is the population of the excited state, $\rho_{gg} = 1 - \rho_{ee}$ is the population of the ground state, and $\rho_{eg}$ is the coherence. Notice that $\Omega$ couples coherences to populations. The effect of decay is twofold: the excited state decays at a rate $\Gamma$, and the coherences decay at $\Gamma/2$. These equations are known as the Optical Bloch Equations. Their steady-state solution is
\begin{align} \label{eq:SSsolnYYZ}
\rho_{ee}^{ss} = \frac{1}{2} \frac{\Omega^2/2}{\Omega^2/2 + \Gamma^2/4 + \delta^2}  
\quad \mbox{and} \quad 
\rho_{eg}^{ss} = \frac{\Omega}{2} \frac{\delta - i \Gamma/2}{\Omega^2/2 + \Gamma^2/4 + \delta^2}.
\end{align}
This steady-state is achieved after several spontaneous emission times. That may be a fast time scale near resonance, and for the strong lines that are typically employed for laser-cooling and imaging \cite{gordonashkin}. However, for the far off-resonant traps used to hold ultracold atoms, the entire experiment of interest may take place before a single photon is scattered. In this case, one should return to the coherent picture presented in Section~\ref{sec:RWAYYZ}. Intermediate cases are challenging, since the effect of ``just a few'' photons depends on the thermalization rate, the motional band structure, and aspects of many-body physics \cite{Daley:2010,gerbiercastin}.

The rate of scattering is $\gamma_{SC} = \Gamma \rho_{ee}$. The atom is a saturable scatterer: the large-intensity limit of the excited fraction is $1/2$, so the maximum steady-state scattering rate is $\Gamma/2$. This saturation of the excited state is a purely quantum effect: due to stimulated emission, a strong drive not only pushes the atom to the excited state, but also pulls it back down to the ground state. On average, the strongly-driven atom spends at most 50\% of its time in the excited state. If we define the saturation intensity $I_{S}$ to be the intensity at which $1/4$ of the population is in the excited state, then we find Eq.~\ref{eq:IsatYYZ} for a two-level system. 

For arbitrary intensity and optical frequency, the scattering rate is
\begin{equation} \label{eq:scattrateYYZ}
\gamma_{SC} = \frac{\Gamma}{2} \frac{\Gamma^2/4 (I/I_{S})}{\delta^2 + \Gamma^2/4 (1 + I/I_{S})}.
\end{equation}
This also shows the power broadening caused by strong intensity: the effective line width is $\Gamma' = \Gamma \sqrt{1 + I/I_{S}}$. Note that for open transitions, this excited-state population is still linear at small intensity, but does not saturate with the same functional behavior.

\subsection{Optical cross section \label{sec:opticalscattYYZ} }

At low power, we can use Eq.~\ref{eq:scattrateYYZ} to find the scattering cross section of a single atom. The flux of photons $\Phi = I / \hbar \omega_L$ is related to the scattering rate (Eq.~\ref{eq:scattrateYYZ}) through the relation $\gamma_{SC} = \sigma_{SC} \Phi$, and thus
\begin{equation} \label{eq:crosssectionYYZ}
\sigma_{SC} = \frac{3 \lambda^2}{2 \pi}\frac{1}{1 + 4\delta^2/\Gamma^2}. 
\end{equation} 
This means that the resonant cross section $3 \lambda^2/2 \pi$ does not depend on the dipole matrix of the resonance. In fact, {\em this is the optical analog of the ``unitarity limit'' } discussed in Chapter~\ref{chapterKokkelmans} and Chapter~\ref{chapterParish} for atom-atom scattering. At a Feshbach scattering resonance, the cross section is $\sigma =\lambda_{T}^2/\pi$, independent of the scattering length. Both resonant cross sections are equal to the square of the wavelength (either optical or de Broglie) times a numerical constant.


\section{Atom-atom interactions \label{sec:interactionsYYZ}}

The inter particle spacing of quantum gases at $n\sim10^{19}$\,m$^{-3}$ is {\em a thousand times larger than the radius} of the electronic clouds around ground-state atoms. This separation of length scales has three major consequences. First, interactions are due to pairwise scattering events. Second, an effective low-energy interaction Hamiltonian can describe the many-body physics of quantum gases to excellent precision (see Chapter~\ref{chapterKollath}). Third, pairwise interactions can be tuned (see Chapter~\ref{chapterKokkelmans}). 

Since atoms have no net charge, their long-range interactions are dipole-dipole. The question is only {\it which} dipole: magnetic or electric.  We start with the effect of induced electric dipole interactions (Section~\ref{sec:vdw}), which is always present and typically dominant. We then discuss permanent dipole interactions (Section~\ref{sec:ddYYZ}), which is a major new topic of study in quantum gases.

\subsection{Short-range interactions \label{sec:vdw} }

As discussed in Section~\ref{sec:StarkYYZ}, ground states of atoms do not have a permanent electric dipole moment. The leading contribution to interactions at long range is a second-order electric dipole-dipole interaction, called the {\em van der Waals} potential, whose magnitude is proportional to $1/R^6$, where $R$ is the internuclear distance. Due to its induced character, the van der Waals potential is isotropic. Further terms have a power-law behavior $1/R^n$ where $n \geq 8$, as discussed in Chapter~\ref{chapterKokkelmans}.

An interaction can be classified as {\it short-range} when the energy depends only on the density of particles. This requires that the interaction decays with a power law $n > D$, for a $D$-dimensional system \cite{Lahaye:2009kf}. Therefore the van der Waals $n \geq 6$ interaction qualifies, and can be treated with an effective contact potential. Electron clouds begin to overlap at sub-nanometer $R$. The difficulty of calculating this hard-core potential (at least for many-electron atoms) defeats quantitative \emph{ab initio} approaches. However, with experimental measurements, the parameters of the effective potential can still be determined precisely, enabling the clear definition of a many-body Hamiltonian valid for the relevant energy scales of ultracold many-body states.

The angular momentum of colliding atoms leads to a centrifugal barrier in the center-of-mass frame of a collision. The two-body wave function can be decomposed into {\em partial waves} with angular momentum eigenvalue $\hbar \ell$, and for which the centrifugal barrier is proportional to $\ell (\ell + 1)$. Because the $\ell = 1$ barrier is on the order of $0.1$\,mK, {\em gases at nanokelvin temperatures can only scatter in the $\ell=0$, or $s$-wave channel}.

The restriction to a single partial wave is a vast simplification of the scattering problem. The result of an elastic collision can only be a phase shift of the two-body wave function. In the limit of low collision energy, the tangent of that phase is simply $-k a_S$, where $k$ is the relative wave vector of the colliding particles and $a_S$ is the \emph{s}-wave scattering length (see Chapter~\ref{chapterKokkelmans}). One can treat the interaction with an effective short-range potential,
\begin{equation} U_\mathrm{eff} = \frac{4 \pi \hbar^2 a_S}{M} \delta(R), \label{eq:UeffYYZ} \end{equation}
where $R$ is the internuclear separation \cite{Dalfovo:1999vn}.

Another consequence of the restriction to \emph{s}-wave collisions is that identical fermions cannot collide. Such a two-body state would be symmetric under particle exchange. Instead, fermions must be in a mixture of two internal (spin) states for cold collisions to occur. In that case, the collision cross section is independent of statistics, and is 
\begin{equation} \sigma = \frac{4 \pi a_S^2}{1 + k^2 a_S^2} \label{eq:csYYZ}. \end{equation}
Bosons in identical internals states have twice this cross section due to symmetrization. Notice that in the limit of $a_S \rightarrow \pm \infty$, this cross section approaches $4 \pi/k^2$, which is the {\em unitary limit} of strong scattering and is independent of the scattering length. Section~\ref{sec:opticalscattYYZ} discusses the optical equivalent of such a unitary limit: in both cases, the cross section is proportional to the square of the wavelength.

\subsection{Long-range interactions \label{sec:ddYYZ}}

Since many neutral atoms have a permanent magnetic dipolar moment in their ground state, one might guess that the strongest interactions between a pair of atoms comes from magnetic dipole-dipole interactions. Such an interaction has a strength $U_{dd} \sim \mu_0 \mu^2/R^3$, where $\mu_0$ is the magnetic constant, and $\mu$ is the magnetic dipole moment shown in Table~\ref{tab:atoms}. 

The phenomenology of this dipolar interaction is radically different from the short-range interaction discussed in the previous section. First, the potential is {\em long-range} in three dimensions, since it does not meet the $n>D$ criterion discussed above. This means that the interaction energy of the system depends not just on density, but also on total number.

Second,  interactions between oriented dipoles are {\em anisotropic}, proportional to $1-3\,\cos^2(\theta)$ where $\theta$ is the angle between the moments. Both attractive and repulsive interactions are possible. The resultant phenomena are discussed in Chapter~\ref{chapterSantos} and Chapter~\ref{chapterPfau}.

However, it turns out that for most atoms, the dipole-dipole interaction is relatively weak, compared to the induced electric dipole interaction discussed in \ref{sec:vdw}. As is explained in Chapter~\ref{chapterPfau}, the magnetic dipole-dipole effects in alkali atoms are roughly $10^2$ smaller than the short-range interaction. 

On the other hand, for elements such as chromium, erbium, and dysprosium, the dipole-dipole interactions are enhanced by $\mu^2$ to be experimentally relevant, or even dominant in some cases (see Table~1 in Chapter~\ref{chapterPfau}). When spin-polarized fermions interact, since \emph{s}-wave collisions are forbidden, thermalization is entirely due to long-range interactions \cite{Lev:161Dy}. Polar molecules held in an optical lattice also have measurable dipole-dipole interactions \cite{Yan:2013ua,Carr:2009wz}.


\section{Creating a quantum gas \label{sec:sequenceYYZ} }

Unlike condensed matter experiments, an ultracold sample is created and destroyed in every experimental cycle. A key component of sample preparation is attaining nanokelvin temperatures, achieved with a succession of two techniques:  laser cooling and evaporative cooling. They take place in a sequence of magnetic and optical traps suited to the energy of the atoms, and to the internal state of the desired quantum gas.

In this section, we describe how the atomic and optical physics introduced so far in this chapter is applied in the 
Toronto $^{40}$K/$^{87}$Rb lattice experiment. For this particular choice of elements, the experimental cycle proceeds similarly in several labs around the world \cite{Roati:2002,Goldwin:2004fw,ETH:2005,Hamburg:2006,Cambridge:2011}.

\subsection{Laser cooling \label{sec:lasercoolingYYZ} }

Laser cooling uses the mechanical effect of light to remove energy and entropy from an ensemble of atoms. Its efficacy relies upon the low entropy of laser light, which is single-frequency, polarized, and often in a single spatial mode. A wide variety of laser cooling techniques have been demonstrated, most of which were invented in the period 1980--2000 \cite{Metcalf,Phillips:1998}, after tunable lasers became a common laboratory tool.

The most widely used cooling technique is {\em Doppler cooling,} in which lasers are tuned to a frequency just below resonance. Atoms are pushed by the recoil momentum $\hbar \mathbf{k}$ of photons scattered at a rate $\gamma_{SC}$ given by Eq.~\ref{eq:scattrateYYZ}. In the rest frame of the atom, each laser beam has a frequency  $\delta - \mathbf{k}\cdot \mathbf{v}$, and thus for $\delta<0$, atoms are ``punished'' for moving towards any incoming laser beam, since that beam is shifted towards resonance ($\delta = 0$). The resultant force is a viscous damping, whose ultimate temperature is $T_D=\hbar\Gamma/2 k_B$, the Doppler temperature \cite{wineland:1979,letokhov:1981,ashkin:1979}.

Laser cooling was first demonstrated using alkali atoms, whose strong transitions typically have $\Gamma/2\pi$ in the range of 5--10\,MHz (see Table~\ref{tab:atoms}). This gives a Doppler temperature of 200\,$\mu$K - already a million times colder than room temperature.
Laser cooling on narrower lines can achieve lower temperatures, as has been demonstrated with earth alkaline atoms. In the case of $^{88}$Sr \cite{vogel:1999,katori:1999}, the broad 30\,MHz cycling transition ($^1$S$_0\rightarrow ^1$P$_1$) at 461\,nm is used to capture atoms, followed by cooling on the narrow 7.5\,kHz forbidden transition ($^1$S$_0\rightarrow ^3$P$_1$) at 689\,nm. This two-step process combines a large capture rate during the first stage with the low Doppler temperature of the second stage. Narrower lines have also been used in alkali atoms  \cite{Duarte:2011,McKay:2011}, but they cannot compare to the kHz-scale lines available in rare earth elements.

Even on a broad line, sub-Doppler temperatures can be achieved using the multi-level structure of ground states \cite{lett:1989,dalibard:1989}. This works especially well for atoms with large $\Delta E_\mathrm{FS}$ (see Table~\ref{tab:atoms}) such as $^{87}$Rb and $^{133}$Cs.

The spontaneous force described above can also be used to {\em trap} atoms, using the multi-level structure of excited states \cite{Chu:1987,Metcalf}. Fortuitously, this works using the same beam powers and polarizations as laser cooling, with a superimposed magnetic quadrupole field. In this magneto-optical trap (MOT), trapping and cooling can proceed simultaneously.

In Toronto, we capture $3\times 10^9$ $^{87}$Rb and $1\times 10^8$ $^{40}$K atoms in a vapor-cell MOT at 200\,$\mu$K, a typical number for these atoms.  It takes about 30\,s to capture the sample. Immediately after the MOT, we have a  30-ms compressed-MOT stage where the $^{87}$Rb and $^{40}$K optical beams are detuned more closely to resonance in order to better mode-match the cloud shape to the magnetic trap. Following the cloud compression, the magnetic fields are switched off momentarily, allowing for sub-Doppler cooling of $^{87}$Rb.

\subsection{Magnetic trapping \label{sec:magtrapYYZ} }

As discussed in Section~\ref{sec:ZeemanYYZ}, atoms must have $m_f g_f > 0$ to be magnetically trapped. A technique called {\em optical pumping} is used to collect atoms in a single ground-state sublevel \cite{Happer:1972uq,OPcohen}. For instance, by scattering light that is circularly polarized along the local field direction, atoms increase their magnetic quantum number $m_f$ by one during an average scattering event. Practically speaking, optical pumping is a quick process: 200\,$\mu$s of our experimental cycle.

The field of a cylindrically symmetric quadrupole trap is $\mathbf{B}(\mathbf{r}) = \beta/2 \mathbf{x} + \beta/2 \mathbf{y} - \beta \mathbf{z}$, where $\beta$ is the on-axis gradient \cite{Ketterle1999}. As trapped atoms move through the field, their magnetic moment follows the direction of $\mathbf{B}$ adiabatically, such that the trapping potential is $m_f g_f \muB |\mathbf{B}|$. However, near the trap minimum, the field approaches zero, and atoms can no longer follow the field direction adiabatically \cite{PETRICH:1995up}. This loss becomes significant at low temperature, so in the final stage of RF evaporation (discussed in Section~\ref{sec:EvaporationYYZ}), we use an optical dipole potential to repel atoms from the center of the trap. This is accomplished with a 760\,nm, 25-$\mu$m-wide beam of roughly 0.5\,W, which creates a 1.5-mK-high barrier to $^{40}$K, and a 0.5-mK-high barrier for $^{87}$Rb. This hybrid trap is called a ``plugged quadrupole trap'' and was used to create the first sodium Bose Einstein condensate (BEC) \cite{Ketterle:23Na}.

Atoms are initially trapped in a magnetic quadrupole field with a gradient $\beta\approx$100\,G/cm. The center of the trap is displaced \cite{Greiner:2001kc} to transport the atoms to a lower-pressure ``lattice chamber'', with improved optical access. The two chambers are separated by a half-meter-long differential tube, such that the pressure is over a thousand times lower in the lattice chamber. The loss of trapped atoms due to collisions with the background gas is proportionally slower. As an alternative to this two-chamber system, one can either load from an atomic beam, or engineer rapid evaporative cooling that can co-exist with a high background vapor \cite{Hansel:2001,Ott:2001,Aubin:chip-DFG}. In the lattice chamber, the magnetic gradient is increased to 230\,G/cm in order to increase the collision rate, for evaporative cooling, discussed in the next section. 

\subsection{Evaporative cooling and sympathetic cooling \label{sec:EvaporationYYZ} }

Evaporative cooling is at work in cooling towers for air conditioning, when we perspire, and when steam rises from a hot cup of tea. Applied to trapped atoms, the high-energy tail of a thermalized cloud will exceed the trap depth, exit the trap, and leave behind a sample with a reduced energy per particle. After further rethermalization, the temperature of the remaining cloud decreases. Unlike in laser cooling, high densities are advantageous for evaporative cooling, and there is no fundamental lower limit to temperature  \cite{Hess:1986,Walraven:1996tk,Luiten:evap,KetterleVanDruten:evap}.

In a successful evaporative cooling, temperature decreases with atom number as $T\propto N^{\alpha}$, where $\alpha$ characterizes the evaporation efficiency.  For instance, for $\alpha=1$, the temperature is reduced by a factor of ten for each factor of ten reduction in atom number. Such a scaling would mean that at least $10^8$ atoms are needed after laser cooling (at approximately $100$\,$\mu$K), if one would like to have $10^5$ atoms left at quantum degenerate temperatures (on the order of 100\,nK - see Section~\ref{sec:introYYZ}).

As the ensemble cools, cloud size decreases as $T^p$ in a D-dimensional trapping potential whose strength is proportional $r^{D/p}$, where $r$ is the distance from the trap minimum \cite{Luiten:evap}. This could increase the density and thus the collision rate $\gamma=n\sigma v_T$; however, atom number decreases during evaporation, and $v_T$ decreases at lower temperature. These three combined effects produce an increasing collision rate for a sufficiently efficient evaporation, such that $\alpha (p-1/2) > 1$  \cite{Walraven:1996tk}. This {\em runaway evaporation} condition is typically a prerequisite for a successful quantum gas experiment. In that case, it is the initial stages of evaporative cooling that are the slowest, motivating continued research on laser cooling techniques to achieve high density at sub-Doppler temperatures.

Buried within the efficiency $\alpha$ are the details of the forced evaporation trajectory, losses due to background collisions, efficiency of energy removal, and the elastic collision rate. At the lowest temperatures, evaporative cooling ceases to be effective either when spatial selection no longer selects the highest energy atoms, or when heat transport is slow. Both of these issues are encountered in optical lattices, for which alternative cooling approaches have been proposed \cite{HoZhou,bernier,Huse:2012}.

The highest-energy atoms will reach the largest magnetic fields during their trajectory, and those that can roll over the maximum potential leave the trap. Evaporation is forced with a RF field that couples trapped $g_{f}m_{f}>0$ states to untrapped $g_{f'}m_{f'}<0$ states. Changing the RF frequency changes the effective edge of the trap. In the plugged quadrupole trap, we start with $5\times 10^{8}$ $^{87}$Rb atoms at 500\,$\mu$K, and sweep an RF field from 50\,MHz to 0.8\,MHz in 25\,s, typically producing a gas of $3\times 10^{6}$ atoms at 10\,$\mu$K.

Although spin mixtures of $^{40}$K can be evaporatively cooled, \cite{Jin:40K} in our apparatus the initial collision rate would be too low to proceed in a reasonable time. Potassium is instead {\em sympathetically cooled} through thermalization with $^{87}$Rb. More generally, sympathetic cooling can be used for a species whose sources are weak or whose laser cooling is challenging. The process relies on elastic collisions between the coolant and target atoms, a way to remove the coolant without removing the target, and a sufficient heat capacity of the coolant. Chapter~\ref{chapterKohl} discusses the  sympathetic cooling of ions with neutral atoms. In our case, we find that inelastic losses are minimized with $^{87}$Rb atoms in the absolute ground state: $f=1$, $m_f=1$. This state is not magnetically trapped, and thus sympathetic cooling is finished in an optical trap.

\subsection{Optical trapping \label{sec:OpticalTrapYYZ}}

A recipe to create a {\em conservative} optical trapping potential is to use very large detuning \cite{Heinzen:1993,Thomas:1999}. Comparing the dipole potential Eq.~\ref{eq:VdipRes} and the scattering rate Eq.~\ref{eq:crosssectionYYZ}, we see that for large detuning the scattering cross section drops faster ($\sim 1/\delta^2$) than the dipole potential ($\sim 1/\delta$). For instance, it is common to use a trapping wavelength several hundred nanometers away from any strong resonance, which is roughly $|\delta|/\Gamma = 10^7$.

Still, there is some resonant enhancement compared to the effect of a static field. For example, for the ``plug'' beam at $760$\,nm mentioned in Section~\ref{sec:magtrapYYZ}, the resonant enhancement is roughly $40$. This is also below the resonances of both $^{40}$K and $^{87}$Rb, creating a {\em repulsive} potential, which is not possible for a static field (see Eq.~\ref{eq:StarkYYZ}).

From the magnetic trap, we transfer atoms into a crossed-beam optical trap at 1054\,nm. This resonant enhancement is roughly $2$, compared to a static field of equivalent strength. A 5-W beam focused to a waist of $50$\,$\mu$m creates a trap depth of 200\,$\mu$K. This is more than sufficient to contain the 10-$\mu$K cloud produced by evaporative cooling in the magnetic trap. Furthermore, since a focused optical beam creates a strong electric field without electrodes, the field can be orders of magnitude greater than the (typically $10^5$\,V/m) discharge limit of an electrode.

In fact, since the depths of optical traps can be greater than laser-cooling temperatures, one can load atoms into an optical trap directly from a laser-cooled cloud \cite{Chapman:2001,Thomas:2000}. On the other hand the trap {\em volume} is small, so more typically an intermediate stage of magnetic trapping and evaporative cooling is used, as in Toronto, to achieve higher final atom number.

After transfer to the optical trap, evaporative cooling is continued. Lowering the beam power forces evaporation, since atoms with an energy higher than the trap depth escape. We transfer $2\times 10^{6}$ $^{87}$Rb atoms at 5\,$\mu$K into a crossed-beam optical trap. After 20\,s, we produce a Bose condensate without any discernible thermal diffraction. By sympathetic cooling in the optical trap, we also produce a quantum degenerate cloud of $^{40}$K atoms.

\begin{figure}[tb!]
\begin{center}
\includegraphics[width=0.6\textwidth]{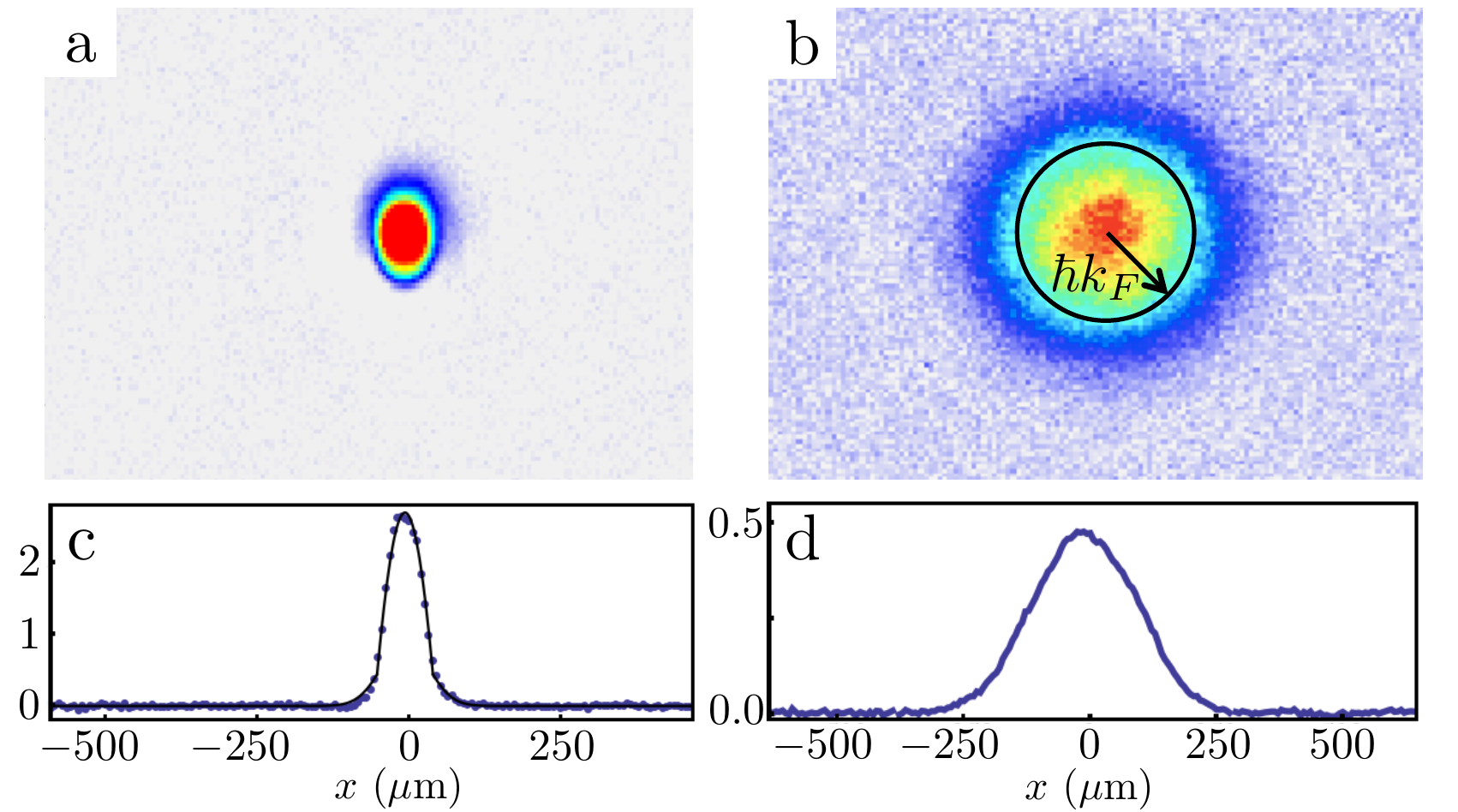} 
\caption{{\bf A degenerate cloud of $^{87}$Rb and $^{40}$K.}  {\bf (a),(c)} A quasi-pure BEC of $5\times10^4$ $^{87}$Rb atoms after evaporation in both a plugged-quadrupole magnetic trap and a crossed-dipole trap, 15\,ms after release from the trap. {\bf (b),(d)} A quantum degenerate cloud of $1.1\times10^5$ $^{40}$K atoms at $320$\,nK, 12\,ms after release from the trap. $\hbar k_F$ is the Fermi momentum. The vertical axis in (c) and (d) is optical density (OD), defined in Eq.~\ref{eq:TorontoOD}, and also represented in color in (a) and (b).}
\label{fig:TorontoQuantumatoms}
\end{center}
\end{figure}

\subsection{Imaging \label{sec:imagingYYZ} }

The first images of quantum degenerate gases were obtained by releasing the cloud from the trap, allowing the density to decrease, and then measuring the absorption of a probe beam passing through the cloud. {\it Absorption imaging} uses the Beer Lambert law, that the attenuation of light is a simple exponential function of the {\em column density}. The resultant intensity $I_a\left(x,y\right)$ is recorded with a camera, where we take $z$ to be the optical axis of the probe beam and imaging system. A second image, $I_0\left(x,y\right)$, without the atoms, is taken to calibrate the intensity of light incident on the cloud. The divided image can be related to the atomic density via the scattering cross section $\sigma_{SC}$ through
\begin{equation} \label{eq:TorontoOD}
-\ln{\left(\frac{I_a\left(x,y\right)}{I_0\left(x,y\right)}\right)}=\sigma_{SC} \int n\left(x,y,z\right)dz  \equiv \mbox{OD} \end{equation}
where $\sigma_{SC}$ is defined in  Eq.~\ref{eq:crosssectionYYZ}. This measured quantity is the {\em optical density} (OD). 

Free-flight expansion before absorption imaging is typically required to reduce the OD to a measurable level. Since the resonant cross section is roughly $\lambda^2$, a cloud with the typical density $n \sim 10^{19}$\,m$^{-3}$ would have an optical attenuation length of $1/n \sigma_{SC} \sim 100$\,nm. This is on the order of the inter particle spacing, and thus only a cloud that is one atom thick could be imaged with resonant absorption! In fact, such an approach is described in Chapter~\ref{chapterChin} to study 2D clouds. However, for typical 3D clouds of $10^5$ atoms, the average radius is several microns, which would give $\mathrm{OD}>10$ in the trap. Instead, the cloud is released, allowed to expand, and imaged. The optimal signal-to-noise ratio is found at OD near unity \cite{Ketterle1999}. An alternative way to reduce OD is to use a high-intensity probe (see Chapter~\ref{chapterChin}).

For ballistic expansion, the acquired image is a convolution of the initial position and velocity distributions of the atoms. This convolution has its simplest interpretation in the long-time limit, when the initial position becomes irrelevant, and one observes the velocity (or single-particle momentum) distribution. This is similar in spirit to the ``far-field'' limit of optical diffraction. For interacting gases, however, the expansion has significant corrections due to interactions, and the imaged distribution is not a Fourier transform of the initial spatial wave function. For Bose condensates, the in-trap distribution is rescaled during expansion, without changing shape \cite{Dalfovo:1999vn,Pethick}.

Figure~\ref{fig:TorontoQuantumatoms} shows absorption images of quantum gases. The different nature of Bose and Fermi statistics is evident when comparing images of $^{87}$Rb and $^{40}$K: the bosonic $^{87}$Rb cloud expands less than the fermonic $^{40}$K cloud.  Whereas bosons ``condense'' into low-momenta states of the trap, Fermi pressure forces fermions apart and into higher-momenta states. In order to gain quantitative information, the time-of-flight distribution is fit to quantum statistical functions  \cite{Ketterle1999,Ketterle2008}.

There are two common alternatives to absorption imaging. Fluorescent imaging can produce a strong signal when the collection angle of the imaging system is high (see Chapter~\ref{chapterWeitenberg}). Alternatively, one can also measure the dispersive effect of atoms on probe light when the probe is not on resonance \cite{KetterlePCI,Bradley:1997tm}.  Both of these techniques are used for \emph{in situ} images.

\begin{figure}[tb!]
\begin{center}
\includegraphics[width=0.6\textwidth]{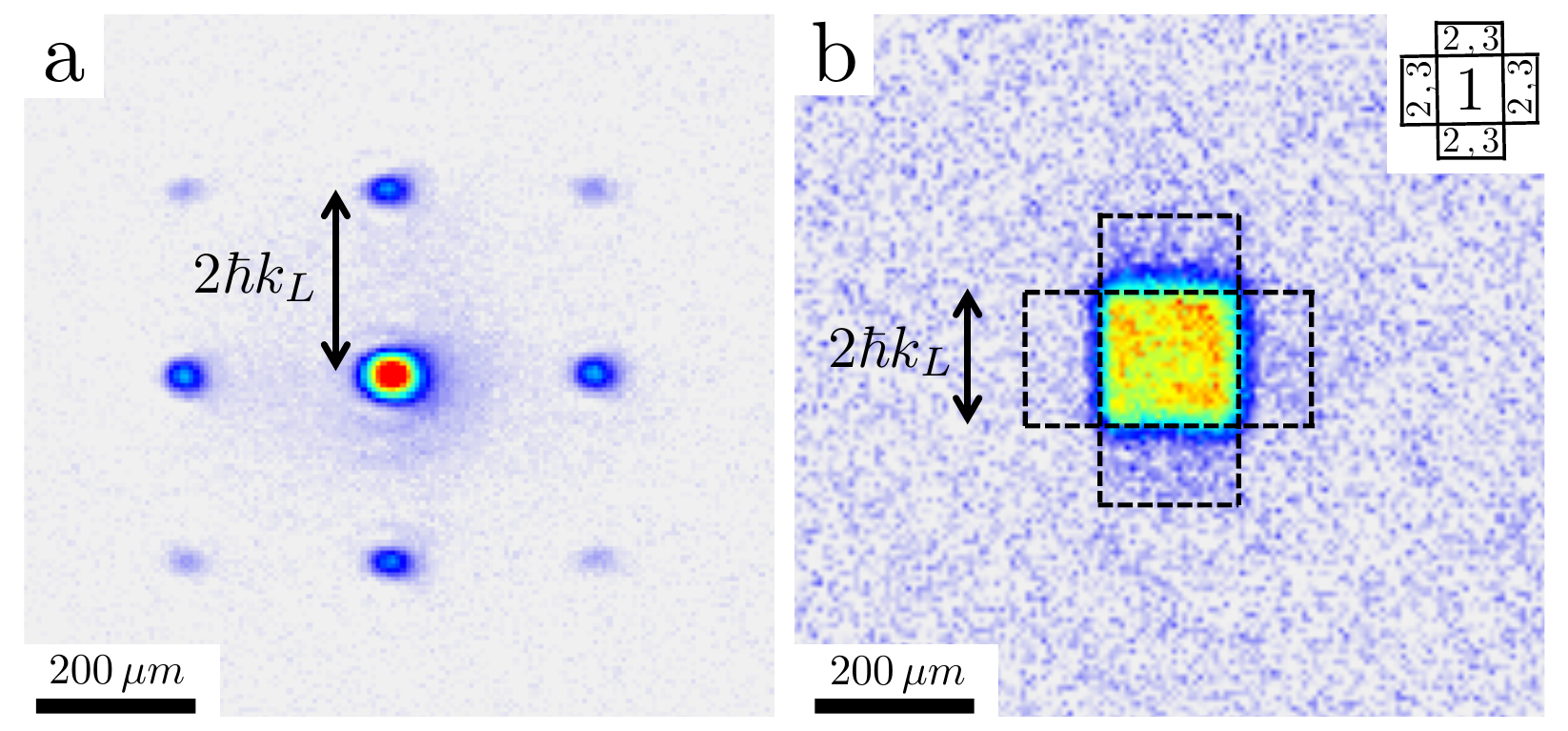} 
\caption{{\bf $^{87}$Rb and $^{40}$K released from a cubic lattice.}  {\bf (a)} Left, a diffracted cloud of $6\times10^4$ $^{87}$Rb atoms after an adiabatic ramp-on and sudden switch-off of a 10$E_R$ lattice, where $E_R=\hbar^2 k_L^2/2m$ and $k_L = 2 \pi/\lambda_L$.  {\bf (b)} Right, a band-mapped cloud of $6\times10^4$ $^{40}$K atoms after an adiabatic ramp-on and fast (200$\mu$s) exponential ramp-off of a 100$E_R$ lattice.  The mask that is super imposed over the image identifies the first three Brillouin zones for a 2D lattice.  The inset figure labels the Brillouin zones and indicates that the second and third Brillouin zones are degenerate in energy.}
 \label{fig:latticeatoms}
\end{center}
\end{figure}

\section{Conclusion: Many-body state preparation and probing}

The conclusion of this chapter is the point of departure for the remaining chapters in this book. Having produced an ultracold sample, one is ready to prepare and probe unique many-body states.

Preparation of interesting many-body states include loading the gas into an optical lattice (Chapter~\ref{chapterKollath}, Chapter~\ref{chapterSengstock}), tuning interactions (Chapter~\ref{chapterKokkelmans}) to the crossover regime (Chapter~\ref{chapterParish}), orienting dipolar moments in the gas (Chapter~\ref{chapterSantos}, Chapter~\ref{chapterPfau}), or photo association of hetero nuclear dimers  \cite{Carr:2009wz}.

Probing states might include RF spectroscopy, Bragg spectroscopy, or modulation spectroscopy (Chapter~\ref{chapterTorma}, Chapter~\ref{chapterTarruell}); or creating currents for transport measurements. The result of these investigations are nearly always learned by imaging, discussed above and in Chapter~\ref{chapterChin} and Chapter~\ref{chapterWeitenberg}. Information is gained both from traditional images and from noise correlations (Chapter~\ref{chapterFolling}).

Figure~\ref{fig:latticeatoms} shows an example of $^{87}$Rb atoms and $^{40}$K atoms being released from a cubic lattice potential in Toronto. In each of three orthogonal directions, counter-propagating beams with $\lambda_L = 1054$\,nm interfere to create a standing wave with a period of $\lambda_L/2$. In Fig.~\ref{fig:latticeatoms}a, the sharp diffraction peaks show phase coherence between bosonic $^{87}$Rb across several sites of the lattice. In Fig.~\ref{fig:latticeatoms}b, the square shape of the momentum expansion shows that $^{40}$K atoms have filled the lowest Brillouin zone of the lattice.

\begin{acknowledgements}
We would like to thank the lattice team (Rhys Anderson, Ryan Day, Graham Edge, and Stefan Trotzky) for their collaboration and contributions to the experiment described in this article. Research was sponsored by NSERC, by AFOSR under agreement number FA9550-13-1-0063, and by ARO with funding from the DARPA OLE program.
\end{acknowledgements}

\bibliography{ChapterThywissen}
\end{document}